\documentclass[aps,prb,groupedaddress,amsmath,amssymb,amsfonts,showpacs,twocolumn]{revtex4}
\usepackage{bm}
\usepackage{graphicx}
\usepackage{color}

\begin{document}

\title{Quantum Charge Fluctuations in a Superconducting Grain}
\author{M. Houzet$^{1,2}$, D.A. Pesin$^3$, A.V. Andreev$^3$, L.I. Glazman$^1$}
\affiliation{
    $^1$ Theoretical Physics Institute, University of Minnesota,
    Minneapolis, Minnesota 55455, USA \\
    $^2$ Commissariat \`a l'\'Energie Atomique, DSM/DRFMC/SPSMS,
    38054 Grenoble, France \\
    $^3$ Department of Physics, University of Washington, Seattle, Washington 98195-1560, USA
}

\begin{abstract}
We consider charge quantization in a small superconducting grain
that is contacted by a normal-metal electrode and is controlled by
a capacitively coupled gate. At zero temperature and zero
conductance $G$ between the grain and the electrode, the charge
$Q$ as a function of the gate voltage $V_g$ changes in steps. The
step height is $e$ if $\Delta<E_c$, where $\Delta$ and $E_c$ are,
respectively, the superconducting gap and the charging energy of
the grain. Quantum charge fluctuations at finite conductance
remove the discontinuity in the dependence of $Q$ on $V_g$ and
lead to a finite step width $\propto G^2\Delta$. The resulting
shape of the Coulomb blockade staircase is of a novel type. The
grain charge is a continuous function of $V_g$ while the
differential capacitance, $dQ/dV_g$, has discontinuities at
certain values of the gate voltage. We determine analytically the
shape of the Coulomb blockade staircase also at non-zero
temperatures.
\end{abstract}


\maketitle

\section{Introduction}

In the conditions of Coulomb blockade, the charge of a grain is well-defined and
discrete. It can be varied by means of the external parameter, gate voltage
$V_g$. Periodically in $V_g$, the ground state of the system approaches a point
of degeneracy, in which two consecutive allowed values of charge yield
the same energy of the ground state. In the limit of no tunneling between the
grain and particle reservoirs (leads), the degeneracy between the ground states
is indeed reached at the corresponding special values of $V_g$. However, electron
tunneling between the grain and leads may remove the degeneracy. The nature of
the resulting many-body ground state is sensitive to the spectrum of
excitations in the grain and in the leads. Except for the case of ultra-small
grains,\cite{vonDelftO1} the level spacing between the single-particle
excitations $\delta$ in metallic grains is usually negligible.  Under this
condition, there are the following three known types of evolution of the ground
state with the variation of the external parameter $V_g$.

If the grain and leads are normal metals (N-I-N junction) the system at the
charge degeneracy point can be mapped on the multi-channel Kondo
problem.\cite{Matveev91} In this mapping, two subsequent values of charge and the
gate voltage play the role of the pseudo-spin and magnetic field, respectively.
(Normal grain charging was experimentally studied in Ref.~\onlinecite{Lehnert}.)
Similar mapping is also possible\cite{Zaikin} for a system consisting of a normal
lead and a superconducting grain (S-I-N junction) in the case of relatively small
charging energy $E_c<\Delta$ ($\Delta$ is the superconducting gap in the spectrum
of excitations); tunneling then lifts the degeneracy between two states with
charges differing by $2e$. The corresponding $2e$ steps in the grain charge were
observed experimentally,\cite{Lafarge93b,Lafarge93a} but there was no detailed
study of the charge dependence on $V_g$ near the degeneracy point.

In the opposite case of a fully superconducting system (S-I-S
junction) with $E_c<\Delta$, the degeneracy removal is equivalent to
the formation of an avoided crossing in a two-level system. Charge
degeneracy for such junctions was studied
experimentally.\cite{Schoelkopf}

The third and the last out of the studied classes of junctions is represented by
a system consisting of a normal grain and a superconducting lead (N-I-S
junction), in which case the degeneracy is not removed by tunneling. Indeed, the
ground state is degenerate with respect to changing the number of electrons in
the grain by one, whereas the particle reservoir may supply only pairs of
electrons.\cite{Matveev1998}

We demonstrate here the existence of a new class of behavior of the
ground state.  It occurs in an S-I-N junction having normal lead
made of a metal and characterized by charging energy $E_c>\Delta$.
An adequate model for such a junction has large number of channels.
We show that in the limit of infinite number of channels, the
dependence of charge $Q$ on $V_g$ has no discontinuities, but the
corresponding differential capacitance, $C_{\text{diff}}=dQ/dV_g$,
remains singular, exhibiting a jump at some value of $V_g$. We find
the full dependence $Q(V_g)$ at a finite (but small, $G\ll 4\pi e^2/\hbar$)
conductance. The smearing of the steps in $Q(V_g)$ due to quantum fluctuations may
be observed in experiments of the type performed in
Refs.~\onlinecite{Lafarge93b} and \onlinecite{Lafarge93a} at higher
values of the junction conductance, or by using the sensitive
charge measuring techniques of Ref.~\onlinecite{Lehnert}. We also
find the evolution of the $Q(V_g)$ dependence with temperature, and
find conditions at which thermal fluctuations do not mask the
quantum effects.

It may seem that within the framework of the constant
interaction model employed here and in the limit of small mean level
spacing in the grain the shape of the Coulomb blockade staircase in
the S-I-N case should not differ from that for the N-I-S system
studied in Ref.~\onlinecite{Matveev1998}. Indeed, the charging
energy of the system, see Eq.~(\ref{eq:1.2}) below, can be
reexpressed in a similar form in terms of the number of electrons in
the lead. Upon this procedure the Hamiltonian of the S-I-N system is
formally identical to that of an N-I-S system. The physical
difference between these cases is that, despite the small mean level
spacing in the grain, for the S-I-N system it is possible in the
experiments to reach the low temperature regime, $T < T^*=\Delta
/\ln(\sqrt{8\pi}\Delta/\delta)$, at which no thermal quasiparticles
are present in the superconductor.~\cite{Tinkhambook} For the N-I-S system
with a macroscopic superconducting lead such regime is beyond
experimental reach because the mean level spacing in the lead,
$\delta_l$, is many orders of magnitude smaller than that in the
grain.

The paper is organized as follows. We present a
simplified derivation of our main results in Section \ref{sec:2}.
In Section \ref{sec:3} we present a rigorous analysis of the
problem, justify our main approximations and evaluate corrections
to them. In Section \ref{sec:4} we present a derivation of our
finite temperature results. In Section \ref{sec:5} we summarize
and discuss our main results.

\section{Qualitative considerations and main results} \label{sec:2}

In the absence of tunneling between a superconducting grain and a
normal lead, the system is described by Hamiltonian
\begin{equation}
\label{eq:1.1} H = H_c+H_{s}+H_{l}\,,
\end{equation}
where $H_c$, $H_{s}$, and $H_{l}$ describe respectively the charging
energy of the grain, including the dependence on the gate voltage,
the BCS state in the grain, and the lead,
\begin{eqnarray}
H_c=E_c({\hat N}-{\cal N})^{2}, && H_{s}=\sum_{k\sigma }\epsilon
_{k}\gamma _{k\sigma }^{\dagger }\gamma _{k\sigma}, \nonumber
\\
H_{l}=\sum_{p\sigma }\xi _{p}c_{p\sigma }^{\dagger }c_{p\sigma}. &&
\label{eq:1.2}
\end{eqnarray}
Here $c^\dagger_{p\sigma}$ and $c_{p\sigma}$ are the creation and annihilation
operators for electrons in the lead, $\gamma_{k\sigma}^\dagger$ and
$\gamma_{k\sigma}$ are the corresponding operators for the Bogolubov
quasiparticles in the superconducting grain; indices $p$ and $k$ denote orbital
states in the lead and grain, respectively, and the spin indices take values
$\sigma=\pm$. We assume that the electron spectrum $\xi_p$ has a constant density
of states near the Fermi level (which is a reasonable assumption for a metallic
lead), while the Bogolubov quasiparticles have a gap $\Delta$ in their spectrum,
$\epsilon_{k}=\sqrt{\zeta_{k}^{2}+\Delta ^{2}}$, where $\zeta_k$ is the electron
spectrum (with the constant density of states, too) in the absence of
superconductivity. The electron number operator in the grain is denoted
by $\hat{N}$. The electrostatic energy of the grain is of the order of
$E_c=e^2/(2C)$ and depends on the gate voltage $V_g$ via the term ${\cal
N}=C_gV_g/e$ in Hamiltonian $H_c$, where $C$ and $C_g$ are the total and gate
capacitances, respectively.

\subsection{position of the step}

In the limit of vanishingly small single particle mean level spacing in
the grain, $\delta \to 0$, the ground state energy of the system is periodic in
gate voltage ${\cal N}$ with period 2. It is also a symmetric function of ${\cal
N}-2l$ with $l$ being an integer. Therefore, when describing the shape of the
Coulomb blockade steps it suffices to study the steps between the charge plateaus
with $2l$ and $2l+1$ electrons in the grain. The shape and position of the $2l-1
\to 2l$ steps is then readily obtained using the aforementioned symmetry
properties.

Without tunneling, the charge of the grain ${\hat Q}=e\hat{N}$ commutes
with the Hamiltonian (\ref{eq:1.1}) and thus is a conserved quantity. Minimizing
the system energy with respect to the discrete number of electrons~$N$, we
find the positions of the steps in charge. The transition between the ``even''
and ``odd'' plateaus with $2l$ and $2l+1$ electrons in the grain, respectively,
occurs at
\begin{equation}
{\cal N}_0=2l+\frac{E_c+\Delta}{2E_c}. \label{eq:1.3}
\end{equation}
In the case of a normal grain ($\Delta=0$), the steps are located at
equally spaced half-integer values of ${\cal N}$. At a finite
$\Delta<E_c$, the ``odd'' plateaus of the Coulomb staircase become
narrower, the positions of the steps being shifted\cite{modphyslett}
by $\pm \Delta/(2E_c)$.

\begin{figure}
\includegraphics[scale=0.8, bb=77 437 323 602]{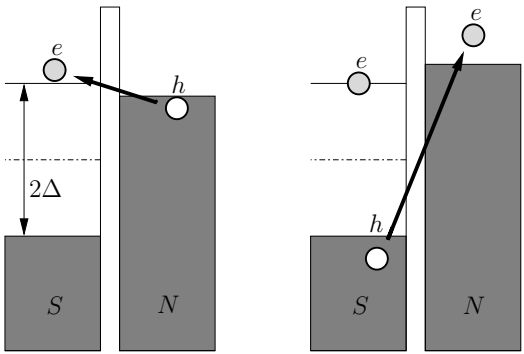}
\caption{\label{fig:1} Picture of the virtual electron-hole pairs
  contributing to the ``vacuum correction'' of the ground state energy
  in a superconducting grain near the charge degeneracy point. On the
  even side of the charge degeneracy points (left panel), without
  coupling, all electrons in the grain are accommodated in the
  condensate. In the presence of the coupling, the virtual states
  consist of a hole created in the lead and an electron in the grain.
  On the odd side (right panel), one quasiparticle is inevitably
  present above the gap and the prevailing virtual states consist in
  creating a hole in the grain and an electron in the lead, with
  energy cost at least $2\Delta$. In comparison, the virtual states
  which result from annihilation of a quasiparticle already present in
  the grain and creating an electron in the lead do not have such
  energy cost. However, by construction the number of such virtual
  states is much smaller than the number of states with the energy gap
  $2\Delta$.  Therefore the former do not contribute significantly to
  the vacuum correction.}
\end{figure}

Tunneling between the lead and the grain results in a correction to the ground
state of the system. The excitation spectra of the superconducting grain and
normal lead differ from each other, leading to the difference in this ``vacuum
correction'' between the even and odd states of the system. To evaluate the
correction, we introduce the tunneling Hamiltonian $H_t$,
\begin{equation}
H_t = \sum_{kp\sigma} t_{kp}a_{k\sigma }^{\dagger
  }c_{p\sigma}+\text{h.c.}
\label{eq:1.4}
\end{equation}
The electron annihilation operators in the grain are related to the Bogolubov
quasiparticle operators by $\gamma _{k\sigma}=u_k a_{k\sigma}-\sigma v_k
a_{-k-\sigma}^{\dagger}$, where $-k$ labels the time-reversed state of $k$, and
$u_k,v_k=(1/\sqrt{2})(1\pm\zeta_{k}/\epsilon_{k})^{1/2}$. The tunnel matrix
elements $t_{kp}$ are related to the conductance of the junction,
\begin{equation}
G=\frac{e^2}{2\pi\hbar}g,\qquad g=8 \pi^2 \sum_{kp} \left| t_{kp}
\right|^{2} \delta ( \zeta_k ) \delta ( \xi_p ). \label{eq:1.5}
\end{equation}
We are interested in the corrections to the ground state energy
introduced by $H_t$ near a degeneracy point. Such ``vacuum
corrections'' appear in the second order of the perturbation in
$H_t$ and result from the formation of virtual electron-hole pairs
across the junction. In the even state, the prevailing (because of
the charging energy) type of pairs has a hole in the normal lead,
while in the odd state holes are predominantly created in the grain.
In the latter state, the creation of a pair involves taking out an
electron from the condensate, which costs extra energy $2\Delta$,
see Fig.~\ref{fig:1}.
The resulting vacuum corrections, $\Delta E_e$ and $\Delta E_o$, at
the even and odd sides of the step, respectively, are different from
each other. This can be easily seen in the second-order perturbation
theory in $H_t$,
\begin{eqnarray}
&&\Delta E_{o}-\Delta E_{e}= \frac{g}{2\pi^2}\int_0^{\infty}
d\xi\int_\Delta^{\infty} d\epsilon \frac{E_c
\epsilon}{\sqrt{\epsilon^2-\Delta^2}} \label{eq:1.6}
\\
&&\times \left(\frac{1}{E_-}+\frac{1}{E_++2E_c}-\frac{1}{E_+}
-\frac{1}{E_-+2E_c} \right ), \nonumber
\end{eqnarray}
where $E_\pm=\epsilon+\xi \pm\Delta$. The corrected
step position, ${\cal N}^*$, to the first order in $g$ is defined by
equation
\begin{equation}
2E_c({\cal N}^*-{\cal N}_0)=\Delta E_o-\Delta E_e.
\end{equation}
In the limit $\Delta \ll E_c$, the result of evaluation of
Eq.~(\ref{eq:1.6}) yields
\begin{equation}
\label{eq:shift} {\cal N}^*={\cal N}_0+\frac{g}{4 \pi^2}
\frac{\Delta}{E_c}\left[1+ \ln \frac{4 E_c}{\Delta}\right] >{\cal
N}_0.
\end{equation}
At $(E_c-\Delta)\ll E_c$, the factor $[1+\ln (4 E_c/\Delta)]$ in
Eq.~(\ref{eq:shift}) should be replaced by
$\sqrt{2}\ln[1/(3-2\sqrt{2})]\simeq 2.49$. Equation (\ref{eq:shift})
should be viewed as the first two terms of the perturbative
expansion for the ${\cal N}^*(g)$ function.

When the conductance of the contact increases, the odd plateaus become shorter,
as it is shown by Eq.~(\ref{eq:shift}). At the same time, the plateaus acquire a
finite slope which can also be calculated in the second-order perturbation theory
in $H_t$ and is of the order of $g/\pi^2$ at integer values of ${\cal N}$. The
slope must be small for the plateaus to be well defined. When $E_c-\Delta \ll E_c$,
odd plateaus are completely suppressed at $g$ exceeding $7.9(1-\Delta/E_c)$.
Depending on the ratio $\Delta/E_c$, this condition can be realized while the even
plateaus still remain flat.

In the following, we neglect the {\it small} correction to the
average grain charge $Q$ related to the finite slope of the
plateaus, and determine the {\it large} (of the order of $1$)
variation of $Q$ which gives the shape of the step separating the
``even'' and ``odd'' plateaus in the $Q$ {\it vs.} $V_g$ dependence.

\subsection{shape of the step at zero temperature}

The shape of the step is described by the dependence of the average
grain charge, $Q=e\langle {\hat N}\rangle$, on the gate voltage. At
zero temperature, it can be found\cite{modphyslett} by
differentiating the ground state energy of the system,
\begin{equation}
\label{eq:def-charge} \langle {\hat N}\rangle_0 = {\cal
N}-\frac{1}{2E_c}\frac{\partial E_{g}}{\partial {\cal N}}.
\end{equation}
The shift of the step position evaluated above comes from the grain
charge fluctuations with a typical energy of the order of $\Delta$.
On the other hand, the shape of the step is determined by the
low-energy excitations. The band width for these excitations is
controlled by the closeness to the (shifted) charge degeneracy
point. This separation of energy scales allows us to derive an
effective low-energy Hamiltonian, $H_0$, in which the
renormalization of the step position is already accounted for. The
new Hamiltonian $H_0$ acts in a narrow energy band of width
$D\ll\Delta$, and is designed to describe the low-energy physics of
the system near the charge degeneracy point,
\begin{equation}
2E_c|{\cal N}-{\cal N}^*|\lesssim D.
\end{equation}
The bandwidth is defined with the energy reference in Hamiltonian
$H_0$ set as the energy of the odd state obtained from Hamiltonian
$H+H_t$ in second order perturbation theory. That is, energy
$E_c(2l+1-{\cal N})^2+\Delta+\Delta E_o$ is subtracted from the
initial Hamiltonian. All the states whose number of electrons on the
grain differs from $2l$ or $2l+1$ have energy higher than $2E_c$.
Thus, they are excluded from the low-energy subspace. Moreover, as a
consequence of $D\ll\Delta $, the even states cannot accommodate any
excitation in the grain (it would cost at least the energy $2\Delta$),
while the odd states accommodate exactly \textit{one} excitation.
Starting from Eqs.~(\ref{eq:1.1}) and (\ref{eq:1.4}) and using the
Schrieffer-Wolff transformation, we are able to derive the low-energy
Hamiltonian, see Section~\ref{sec:3}. Here we present only the
relevant for the discussion terms of that Hamiltonian:
\begin{eqnarray}
H_0&=&\epsilon_0 \left|0\right>\left<0\right| + \sum_{k\sigma}
\frac{\zeta^2_k}{2 \Delta} \left|k\sigma\right>\left<k\sigma\right|
+\sum_{p\sigma }\xi _{p}c_{p\sigma }^{\dagger }c_{p\sigma} \nonumber
\\
 & & + \sum_{kp\sigma} \left[ \tilde{t}_{kp}\left|k\sigma\right>\left<0\right| c_{p\sigma}  +
\tilde{t}_{kp}^* c^{\dagger}_{p\sigma}
\left|0\right>\left<k\sigma\right| \right],
\label{eq:effective-hamiltonian}
\end{eqnarray}
Here $\epsilon_0$ is the energy of the grain in the ``even'' state
$|0\rangle$; it accounts for the renormalization of the step
position by virtual electron-hole pairs with energy exceeding $D$.
Up to small terms of the order of $g\sqrt{\Delta D}$, this energy is
$\epsilon_0=2E_c({\cal N}-{\cal N}^*)$, cf. Eq.~(\ref{eq:shift}).
The energies of allowed states for the Hamiltonian
Eq.~(\ref{eq:effective-hamiltonian}) are within the band of width
$\sim 2D$. States $\left|k\sigma\right>$ have an excitation in state
$(k\sigma)$ of the grain with energy $\epsilon_k-\Delta \approx
\zeta^2_k/(2 \Delta)$. The tunnel matrix elements
$\tilde{t}_{kp}=u_kt_{kp}\approx (1/\sqrt{2})t_{kp}$ account for the
coherence factors values at small energies. Note, that the
Hamiltonian (\ref{eq:effective-hamiltonian}) allows only for zero or
one additional electron in the grain.

With the change of energy reference, Eq.~(\ref{eq:def-charge})
defining the average grain charge must be replaced by
\begin{equation}
    \label{eq:def-charge-eff}
    \langle {\hat N}\rangle_0 = 2 l + 1-\frac{1}{2E_c}\frac{\partial
    E_{0}}{\partial {\cal N}},
\end{equation}
where $E_0$ is the ground state energy of Hamiltonian
(\ref{eq:effective-hamiltonian}).

In the zeroth order in $\tilde{t}_{kp}$, the wave function of an
even state is a direct product of $|0\rangle$ for the grain and some
state $|f\rangle$ of the Fermi sea in the lead; the ground state is
$|0\rangle\otimes |f_0\rangle$. Tunneling terms in
Eq.~(\ref{eq:effective-hamiltonian}) modify the eigenfunctions. In
the lowest order of perturbation theory, the wave function acquires
the form
\begin{equation}
\label{eq:wavefunction}
\left|\Psi\right>=\left(A\left|0\right>+\sum_{kq\sigma}\beta_{kq\sigma}
c_{p\sigma}\left|k\sigma\right>\right)\otimes \left|f\right>.
\end{equation}
The second term in parentheses in Eq.~(\ref{eq:wavefunction})
describes amplitude of a quasiparticle-hole pair state created due
to the tunneling; this amplitude is small within the perturbation
theory. In higher orders of the perturbation theory, additional
electron-hole pairs may be created in the normal lead. However, as
we show in the next Section, these additional terms are small if the
number $N_{\rm
  ch}$ of quantum channels in the junction is large. In the simplest
case, the number of channels is of the order of the junction area
measured in units of the Fermi wavelength in the metallic electrodes
(see Sec.~\ref{sec:3} for details), and the condition $N_{\rm ch}\gg
1$ is not restrictive. If it is satisfied, the wavefunction
Eq.~(\ref{eq:wavefunction}) is valid beyond the perturbation theory.
Having this in mind, for now we may use it as a trial function for
an eigenstate originating from the state $|f\rangle$ in the absence
of tunneling. Then the ground state energy is the lowest value of
$E_0$ which solves the equation
\begin{equation}
E = \epsilon_0 +
    \sum_{kp\sigma}
    \frac
        {\left|\tilde{t}_{kp}\right|^2 \theta (-\xi_p )}
        {E-\zeta^2_k/(2\Delta) +\xi_p}.
\label{eq:equation-ground-state}
\end{equation}
This lowest-energy solution
\begin{eqnarray}
   E_0({\cal N})
   &=&
   -2E_c \delta {\cal N}
   \left(
        \sqrt{1+ \frac{ {\cal N}^*-{\cal N}}{\delta {\cal N}}}
        -1
   \right)^2
   \theta({\cal N}^*-{\cal N}),
   \nonumber
   \\
   \delta{\cal
   N}&=&\left(\frac{g}{8\pi}\right)^2\frac{\Delta}{E_c},
   \label{eq:ground-state-energy}
\end{eqnarray}
is separated at ${\cal N}<{\cal N}^*$ from the continuum of states
$E>0$ allowed by Eq.~(\ref{eq:equation-ground-state}) and can be
associated with a bound state. This state is formed by the
quasiparticle (virtually) populating the grain and the corresponding
hole in the lead. Upon approaching the threshold value ${\cal N}={\cal
  N}^*$, the binding energy of this state vanishes, $E_0({\cal
  N})\propto ({\cal N}-{\cal N}^*)^2$. At ${\cal N}\geq {\cal N}^*$
the hole is not localized near the junction any more; the formerly
discrete energy level merges with the edge of the continuum
spectrum.  The described qualitative change of the spectrum at the
even-odd transition is identical to the one in a well-known problem of
{\it single-particle} quantum mechanics. Indeed, the same
transformation occurs with the spectrum of a particle attracted to a
three-dimensional well upon the gradual reduction of the well depth;
particle becomes delocalized at a certain strength of the
potential,\cite{Fano} and the discrete energy level ceases to exist.

\begin{figure}
\includegraphics[scale=0.8, bb=129 484 335 623]{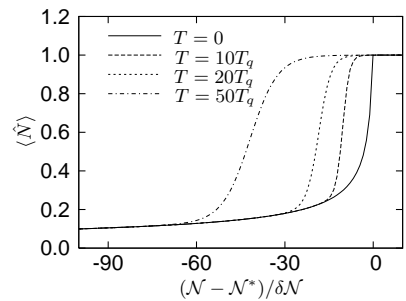}
\caption{\label{fig:2} Coulomb staircase at different temperatures
$T=0$ (thick line) and $T=10,20,50 T_q$. $\langle {\hat
N}\rangle$ is the average number of electrons in the grain, $V_g$
is the gate voltage. The plots correspond to the following
values of parameters: $g=2$, $\Delta/E_c=0.7$, $\sqrt{8
\pi}\Delta/\delta=10^4$. With these parameters, $T^*\approx
0.1\Delta$, $T_q\approx 0.001\Delta$, and ${\cal
N}^*\approx2l+0.8$. The correction due to the finite slope of the
steps at integer values of $C_gV_g/e$ is not accounted for in the
plots.}
\end{figure}

The number of electrons in the grain at zero temperature is obtained
from Eqs.~(\ref{eq:def-charge-eff}) and
(\ref{eq:ground-state-energy}):
\begin{eqnarray}
\langle {\hat N}\rangle_0 &=& 2l+1 -\theta ({\cal N}^*-{\cal N})
f\left(\frac{{\cal N}^*-{\cal N}}{\delta {\cal N}}\right)\,,
\nonumber\\
f(x) &=& 1-\frac{1}{\sqrt{1+x}}.
    \label{eq:ground-state-charge}
\end{eqnarray}
Eq.~(\ref{eq:ground-state-charge}) describes the transition between
the ``even'' plateau with $2l$ electrons in the grain and the
``odd'' plateau with $2l+1$ electrons. The dependence of $\langle
{\hat N}\rangle_0$ on ${\cal N}$ is shown in Fig.~\ref{fig:2}.
At the degeneracy point (${\cal N}={\cal N}^*$), the charge is a
continuous function of ${\cal N}$, but there is a jump in the
differential capacitance, $C_{\text{diff}}=C_g d\langle {\hat
N}\rangle_0/d{\cal N}$. We can define the step position by equation
$\langle {\hat N}\rangle_0=2 l +\frac{1}{2}$ and characterize the
step width by the value of $W_0=(d\langle {\hat N}\rangle_0/d{\cal
N})^{-1}$ at the step position.  The smearing of the steps of the
Coulomb staircase occurs in the second-order in $g$, with a typical
width $W_0=16 \delta{\cal N}$.

\subsection{shape of the step at finite temperature}

We turn now to the effect of a finite temperature, $T$, on the
average grain charge, $\langle{\hat N}\rangle$.
Equation~(\ref{eq:def-charge-eff}) can still be used, except that
$E_0$ should be replaced with the free energy $F=-T\ln Z$, where
\begin{equation}
    \label{eq:Z-def}
    Z=\mathrm{Tr}\exp\left(-\beta H_0 \right)
\end{equation}
is the partition function of the system\cite{modphyslett} described
by Hamiltonian (\ref{eq:effective-hamiltonian}).

To compute $Z$, we must determine the energies of the excited
states. For that we once again will be using
Eq.~(\ref{eq:wavefunction}) as a trial wave function. The
corresponding eigenenergy can be represented in the form $E={\cal
E}_f+{\cal E}$. Here ${\cal E}_f$ is the energy of the ``bare''
state $|f\rangle$. In the limit of a thermodynamically
large lead, ${\cal E}$ is a solution of the already used
Eq.~(\ref{eq:equation-ground-state}), where the step
function $\theta(\xi)$ should be replaced by the Fermi
distribution function. Similarly to the zero temperature case
there are two classes of solutions at ${\cal N}<{\cal N}^*$. In
the first class there is just one discrete solution whose energy
at low temperatures is ${\cal E}=E_0$, see discussion in
the paragraph below Eq.~(\ref{eq:Z}). The second class is
represented by continuum spectrum ${\cal
E}=\zeta_k^2/(2\Delta)-\xi_p$. The full energy for a state in the
second class can be written in the form $E={\cal
\bar{E}}_f+\zeta^2/(2\Delta)$, where ${\cal \bar{E}}_f$ is
the energy of the state that differs from $|f\rangle$ by the
presence of one hole in state $p$. Since the partition function
involves summation over all states the difference between ${\cal
\bar{E}}_f$ and ${\cal E}_f$ drops out and we have,
\begin{equation}\label{eq:Z_approx}
Z=\sum_fe^{-\beta({\cal E}_f+E_0)}+\sum_{f,k\sigma}e^{-\beta({\cal
E}_f+\zeta^2_k/(2\Delta))}.
\end{equation}
Factorizing the ${\cal N}$-independent partition function of the
lead, $Z_\text{lead}=\sum_fe^{-\beta{\cal E}_f}$, and integrating
the second term over $k$, we get
\begin{equation}
    \label{eq:Z}
    Z=Z_\text{lead}\left[e^{-\beta E_0}+N_\text{eff}(T)\right],
    \quad
    N_\text{eff}(T)=\sqrt{\frac{8 \pi \Delta T}{\delta^2}},
\end{equation}
where $\delta$ is the one-electron level spacing in the grain. A
rigorous derivation of the finite-temperature partition function
is detailed in Section~\ref{sec:4}.

We would like to note that the replacement of
$\theta(\xi)$ by the Fermi function in
Eq.~(\ref{eq:equation-ground-state}) leads to a small difference
between the energy of the discrete state and $E_0$. This finite
temperature correction is small in $T/E_0$. For the purpose of
evaluating the partition function this correction can be ignored
at all temperatures because at $T/E_0 \gtrsim 1/\ln
N_\text{eff}(T)$, see Eq.~(\ref{eq:Z}), the partition function
is already dominated by the contribution from the states of the
continuum, while $T/E_0$ is still small.

Using Eq.~(\ref{eq:def-charge-eff}), we obtain the average grain
charge at finite temperature
\begin{equation}
\label{eq:charge} \langle{\hat N}\rangle =2l+1- \frac{\theta ({\cal
N}^*-{\cal N})} {1 +  N_\text{eff}(T) e^{\beta E_0({\cal N})}}
f\left(\frac{{\cal N}^*-{\cal N}}{\delta {\cal N}}\right),
\end{equation}
where $E_0({\cal N})$ and $f(x)$ are defined in
Eqs.~(\ref{eq:ground-state-energy}) and
(\ref{eq:ground-state-charge}), respectively.

\begin{figure}
\includegraphics[scale=1, bb=212 598 416 737]{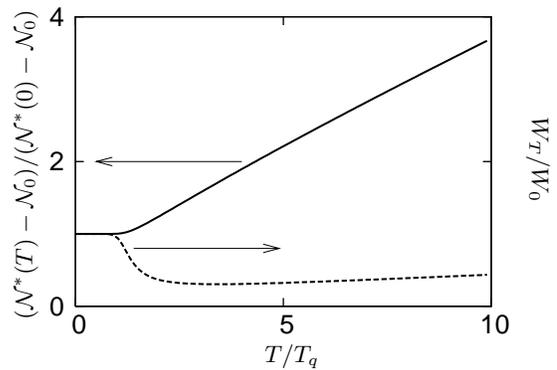}
\caption{\label{fig:3} Temperature dependence of the step
position, ${\cal N}^*(T)$, and width $W_T$. The same values of
parameters as in fig.~\ref{fig:2} is used.}
\end{figure}

The temperature dependence of the Coulomb staircase is shown in
Fig.~\ref{fig:2}. At $T\to 0$ Eq.~(\ref{eq:charge}) reproduces the
average charge, $\langle{\hat N}\rangle_0$, given by
Eq.~(\ref{eq:ground-state-charge}). At finite temperature, the odd
plateaus become broader. The position of the step, defined by
equation $\langle{\hat N}\rangle=2l+\frac{1}{2}$, results from the
competition between the energy of the grain in an even state, $E_0$,
and the entropy of the large number of odd states, given by
$N_\text{eff}(T)$ in Eq.~(\ref{eq:Z}). The step is shifted from
${\cal N}={\cal N}^*$ to ${\cal N}={\cal N}^*(T)$,
\begin{equation}
\label{eq:thermal-shift} {\cal N}^*(T)={\cal N}^*-\frac{T}{2E_c}\ln
N_\text{eff}(T)
\end{equation}
when
\begin{equation}\label{eq:T_q}
T \agt T_q = \left(\frac{g}{8 \pi}\right)^2
\frac{\Delta}{\ln\frac{g\Delta}{\delta\sqrt{8\pi}}},
\end{equation}
that is when the shift exceeds the zero-temperature width of the
step, $W_0=16\delta{\cal N}$. Note that the thermal width
$W_T=T/E_c$ of the charge step at $T \gtrsim T_q$ is smaller than that at zero
temperature, $W_{T_q}\lesssim W_0$. We thus expect a strong
nonmonotonic temperature dependence of the step width with
the minimum width occurring at $T\sim T_q$. The temperature
dependence of the step position and width are shown in
Fig.~\ref{fig:3}.

The junction conductance $g$ does not affect the charge steps at
$T\gg T_q$. At even larger temperature,
$T\sim T^* =\Delta/\ln(\sqrt{8  \pi}\Delta/\delta)$,
the ``odd" plateaus were shown to reach the same size as the
``even" ones.\cite{modphyslett}

In the rest of the paper, we aim at giving a rigorous derivation of
the main results, Eqs.~(\ref{eq:ground-state-energy}),
(\ref{eq:ground-state-charge}), and (\ref{eq:charge}) . We'll show
that they hold for the experimentally relevant case of a wide
multichannel junction between the grain and the lead.  In Section
\ref{sec:3}, we derive the low-energy effective Hamiltonian
(\ref{eq:effective-hamiltonian}); we demonstrate there that the
corrections in the number of channels to the ground state energy,
are small.  In Section \ref{sec:4}, we derive the partition function
for our system.

\section{Effective Hamiltonian and the grain charge at zero temperature} \label{sec:3}

In this Section we derive the effective low-energy Hamiltonian
(\ref{eq:effective-hamiltonian}) which is used to describe the shape
of the step of the Coulomb staircase between the even state of the
grain with a charge $Q=(2l)e$ and the odd state with a charge
$Q=(2l+1)e$. We show that it gives the same physics as the
Hamiltonian of the system, $H+H_t$, given by Eqs.~(\ref{eq:1.1}) and
(\ref{eq:1.4}), for a wide multichannel contact between the grain
and the lead.

The effective Hamiltonian $H_0$ is acting in a narrow bandwidth $D$,
on a limited set of states which was characterized in Section
\ref{sec:2}: only low-energy electrons (or holes) may be excited in
the lead, and the states of the grain are only those with zero or
one low-energy excitation,
\begin{equation*}
|\xi_k|,E_c({\cal N}^*-{\cal N})+\zeta^2/2\Delta\lesssim D.
\end{equation*}
The states outside the bandwidth $D$ are accounted in
$H_0$ perturbatively. For the present problem, it is enough to
calculate such contribution in the second order in $H_t$. Therefore,
the appropriate Schrieffer-Wolff transformation of the initial
Hamiltonian is \cite{QM}
\begin{equation}\label{eq:RG}
H_0\approx M(H+H_t)M-MH_t(1-M)\frac{1}{H}H_tM.
\end{equation}
Here, $M$ is the projector operator on the unperturbed states whose
energy lies within the band $D$. We obtain
\begin{eqnarray}
H_0 && =H'+V, \label{eq:effective-hamiltonian-2}
\\
H' && = \epsilon_0 P_e  + \sum_{k\sigma} \frac{\zeta^2_k}{2\Delta}
P_o \gamma^\dagger_{k\sigma} \gamma_{k\sigma} P_o
+\sum_{p\sigma}\xi_pc^\dagger_{p\sigma}c_{p\sigma} \nonumber
\\
V & & = \sum_{kp\sigma} \left[ \tilde{t}_{kp}P_o
\gamma^\dagger_{k\sigma} c_{p\sigma} P_e + \tilde{t}_{kp}^*P_e
c^{\dagger}_{p\sigma}\gamma_{k\sigma} P_o \right] \nonumber
\\
& & + \sum_{p\neq p^{\prime}\sigma} \sum_{i=e,o} w_{pp^{\prime}}^i
c_{p\sigma}^{\dagger}c_{p^{\prime }\sigma} P_i + \sum_{k\neq
k^{\prime }\sigma } w_{kk^{\prime }} P_o \gamma^\dagger_{k\sigma}
\gamma_{k'\sigma} P_o. \nonumber
\end{eqnarray}
Here, $P_e$ and $P_o$ are projection operators on the even and odd
states with $2l$ or $2l+1$ electrons in the grain, respectively,
$\tilde{t}_{kp}=u_kt_{kp}$, and
\begin{equation}
\epsilon_0 =2E_c({\cal N}_0-{\cal N}^*) +\sum_{kp\sigma}'
    \frac
        {\left|\tilde{t}_{kp}\right|^2 \theta (-\xi_p )}
        {\frac{\zeta^2_k}{2\Delta} -\xi_p}.
\label{eq:ug}
\end{equation}
The prime here means that only excited states in the bandwidth $D$,
(i.e., such as $|\zeta^2_k/(2\Delta) -\xi_p| <D$) are included.
Making the
summation on $k$ and $p$, we obtain
\begin{equation} \label{eq:omega}
\epsilon_0\approx 2E_c\left({\cal N}-{\cal N}^*+\frac{g}{2\pi^2}
\frac{\sqrt{2\Delta D}}{2E_c}\right).
\end{equation}
The degeneracy point ${\cal N}$ defined by condition $\epsilon_0=0$
would slightly differ from ${\cal N}^*$. This difference comes from
the virtual states with energies within the band $D$; such states
were accounted for in the evaluation of ${\cal N}^*$, but do not
contribute to $\epsilon_0$. This difference is small at small
($D\ll\Delta$) bandwidth. Note, that by construction of the
Hamiltonian $H_0$, we need energy $\epsilon_0$ at ${\cal N}={\cal
N}^*$ to be within the band $D$. This sets a condition $D\gtrsim
\frac{g^2}{2\pi^4}\Delta$, which does not contradict our initial assumption $D\ll
\Delta$.

Perturbation $V$ in Eq.~(\ref{eq:effective-hamiltonian-2}) has terms
describing electron-hole pair fluctuations across the junction, and
terms describing scattering off the junction within the grain or
lead. The latter terms correspond to second-order processes in $H_t$
generated by the Schrieffer-Wolf transformation (\ref{eq:RG}). For
instance, the term corresponding to scattering off the lead when the
grain is in the even state has probability amplitude
\begin{equation}\label{eq:W}
w_{pp'}^e=-\sum''_{k}t_{kp'}t_{kp}^* \left(
\frac{u_k^2}{\epsilon_k-\Delta}-\frac{v_k^2}{2E_c+\epsilon_k+\Delta}
\right).
\end{equation}
Here, the first and second term in parentheses correspond to
virtual excitations with charge  $(2l+1)e$ and $(2l-1)e$ in the
grain, respectively; the double-prime means that only virtual states
with excitation energy outside the bandwidth are included in the
sum, it reduces to condition $\epsilon_k-\Delta>D$ for the first
term (we also used property $t_{kp}^*=t_{-k-p}$ valid for tunneling
Hamiltonian preserving the time-reversal symmetry). The amplitudes
$w^o_{pp'}$ and $w_{kk'}$ have expressions similar to
Eq.~(\ref{eq:W}). We will put terms proportional to $w^i_{pp'}$ and
$w_{kk'}$ in $V$ aside for a while and return to the discussion of
their effect at the end of the Section.

Let us calculate the ground state energy of $H_0$.  At ${\cal N}<{\cal
  N}^*$, the unperturbed ground state on the ``even'' side,
$\left|e\right>=|0\rangle\otimes|f_0\rangle$, is the direct product
of the BCS ground state with $2l$ electrons in the grain,
$|0\rangle$, times Fermi sea ground state in the lead,
$|f_0\rangle$. Its bare energy is $\epsilon_0$. In the presence of
$V$, we determine its renormalized energy, $E_0$, with
Brillouin-Wigner perturbation theory.\cite{QM} It is the solution of
the infinite order equation
\begin{eqnarray}
&&E_0 = \epsilon_0 + \left<e\right|V\left|e\right> +
    \sum_{n\neq 0}
    \frac
        {
            \left<e\right|V\left|n\right>
            \left<n\right|V\left|e\right>
        }
        {E_0-\epsilon_n}
\label{eq:BW}
\\
&&+
    \sum_{n_1,\cdots,n_j\neq 0}
    \frac
        {
            \left<e\right|V\left|n_1\right>
            \left<n_1\right|V\left|n_2\right>
            \cdots
            \left<n_j\right|V\left|e\right>
        }
        {(E_0-\epsilon_{n_1})\cdots(E_0-\epsilon_{n_j})}
+\cdots, \nonumber
\end{eqnarray}
where $\left|n\right>$ are eigenstates of $H'$, with energy
$\epsilon_n$. Anticipating that, for typical S-I-N systems with
multichannel contacts, the series in the Eq.~(\ref{eq:BW}) can be
truncated at the term of the second order in $V$, we obtain:
\begin{equation}
E_0 = \epsilon_0 +
    \sum_{kp\sigma}'
    \frac
        {\left|\tilde{t}_{kp}\right|^2 \theta (-\xi_p )}
        {E_0-\frac{\zeta^2_k}{2\Delta} +\xi_p}.
 \label{eq:selfconst_energy_even}
\end{equation}
Here again, the prime means that only excited states in the
bandwidth $D$ are included. This equation is identical to
Eq.~(\ref{eq:equation-ground-state}). As a result of the summation
in Eq.~(\ref{eq:selfconst_energy_even}), the bandwidth $D$
disappears from the equation for the ground state energy and we
arrive at Eq.~(\ref{eq:ground-state-energy}).

\bigskip

The above result holds when higher order in $V$ terms can be
neglected in Eq.~(\ref{eq:selfconst_energy_even}).  This is the
case for a generic tunnel junction between a lead and metallic
grain. Indeed, typically the area of junction $S$ exceeds
significantly the square of the Fermi wavelength in a metal. The
effective number of quantum channels in the junction,
$N_\text{ch}\sim Sk_F^2\gg 1$ provides us with a parameter
allowing the truncation of series
Eq.~(\ref{eq:selfconst_energy_even}) in the ballistic regime.
In a realistic setup, electrons are backscattered to the
junction from the impurities and the boundaries in the
grain and in the lead. In a typical situation of a small junction to a
macroscopic lead the the backscattering of electrons from the lead
to the junction may be neglected. We therefore concentrate on the
effects of electron returns from the grain to the junction. Due to
the finite grain size such returns are bound to occur. Quantum
interference between returning electron trajectories in
the grain may lead to a reduction of $N_\text{ch}$.\cite{Hekking}
However, we shall see that the corresponding contribution to
Eq.~(\ref{eq:ground-state-charge}) remains small in the parameter
$1/N_{\rm ch}^{\rm eff}=\delta/g\Delta\ll 1$.

We start with the analysis of higher-order perturbation theory
terms which involve matrix elements ${\tilde t}_{kp}$ only. Let us
first consider the ``even" side of the transition, at ${\cal
  N}<{\cal N}^*$. The first term in Eq.~(\ref{eq:BW}) that we neglect
is
\begin{equation}
J=\sum_{kk'pp'\sigma}' \frac
{\theta(-\xi_p)\theta(\xi_{p'})\tilde{t}_{kp}\tilde{t}_{kp'}^*
\tilde{t}_{k'p'}\tilde{t}_{k'p}^*} {[E_0-\epsilon_{kp}]
[E_0-\epsilon_{pp'}] [E_0-\epsilon_{k'p}] }, \label{eq:fourth-order}
\end{equation}
where $\epsilon_{kp}=\zeta^2_k/(2\Delta)-\xi_p$ and
$\epsilon_{p'p}=\epsilon_0+\xi_{p'}-\xi_p$, and the prime means that
$|\epsilon_{kp}|,|\epsilon_{k'p}|,|\epsilon_{p'p}|<D$. We define the
correlation function
\begin{eqnarray}
K(\xi,\xi';\zeta,\zeta')=\sum_{kk'pp'} t_{kp} t_{kp'}^* t_{k'p'}
t_{k'p}^* \delta (\xi-\xi_p )\nonumber
\\
\times \delta (\xi'-\xi_{p'}) \delta (\zeta-\zeta_k ) \delta
(\zeta'-\zeta_{k'} ), \label{eq:correlator}
\end{eqnarray}
which simplifies Eq.~(\ref{eq:fourth-order}) to
\begin{equation}
J=-\int' d\xi d\xi' d\zeta d\zeta' \frac
{\theta(-\xi)\theta(\xi')K(\xi,\xi';\zeta,\zeta')}
{[\epsilon_{\zeta\xi}-E_0] [\epsilon_{\xi'\xi}-E_0]
[\epsilon_{\zeta'\xi}-E_0] },  \label{eq:fourth-order-cont}
\end{equation}
with $\epsilon_{\zeta\xi}=\zeta^2/(2\Delta)-\xi$ and
$\epsilon_{\xi'\xi}=\epsilon_0+\xi'-\xi$; the prime means
$|\epsilon_{\zeta\xi}|,|\epsilon_{\xi'\xi}|,|\epsilon_{\zeta'\xi}|<D$.

The value of $J$ depends on a concrete model of the junction. We
consider here a thin homogeneous insulating layer separating the grain
from lead; the appropriate barrier potential is $U(\mathbf
{r})=U_0\delta(z)$, where $z$ is the distance from the interface.  The
transmission coefficient ${\cal T}$ for such a barrier depends on the
angle of incidence for the incoming electron characterized by the
normal component $k_z$ of its momentum $\mathbf{k}$; in the limit of
low barrier transparency ${\cal T}(k_z)=(k_z/m U_0)^2\ll 1$. The
dimensionless conductance of the junction is $g={\cal T}Sk_F^2/4\pi$,
where $k_F$ is the Fermi wavevector and $S$ is the area of the
junction and ${\cal T}={\cal T}(k_F)$. We can introduce the number of
channels in the junction by expressing the conductance
$g={\overline{\cal T}}N_{\rm ch}$ in terms of the angle-averaged
transmission coefficient, ${\overline{\cal T}}={\cal T}/3$. This
definition of $N_{\rm ch}$ yields $N_{\rm ch}=3Sk_F^2/4\pi$. In terms
of the matrix elements in the tunneling Hamiltonian, the above model
of the barrier corresponds to
\begin{equation} \label{eq:matrix-elements}
t_\mathbf{kp}=\frac{\sqrt{\cal T}}{\nu\sqrt{V_lV}}
k_zp_z\delta^2(\mathbf{k}_\|-\mathbf{p}_\|).
\end{equation}
with $\nu=(mk_F/2\pi^2)$, $V$, and $V_l$ being the density of states,
volume of the grain and volume of the lead, respectively; $m$ is the
effective electron mass.  Equation~(\ref{eq:matrix-elements}) accounts for
the conservation of the component $\mathbf{k}_\|$ of the electron
momentum parallel to the barrier. Inserting now
Eq.~(\ref{eq:matrix-elements}) into (\ref{eq:correlator}), we obtain
the correlation function in the ballistic regime
\begin{equation}
K_\text{b}(\xi,\xi';\zeta,\zeta')=\frac{3 g^2}{8 \pi^4 N_\text{ch}}.
\end{equation}
We can now evaluate Eq.~(\ref{eq:fourth-order-cont}). It yields the
dominant contribution in $|E_0|$:
\begin{equation}
J_\text{b}=-\frac{g^2}{8 \pi^2 N_\text{ch}}\Delta \ln
\frac{D}{-E_0}\ln \frac{D}{\epsilon_0}. \label{eq:ballistic_result}
\end{equation}
In the vicinity of the degeneracy point, at ${\cal N}^*-{\cal
  N}\alt\delta{\cal N}$, the energy $E_{0}$ is much smaller than
$\epsilon_0 \propto g(\Delta D)^{1/2}$, as the bandwidth $D$ satisfies
the conditions $\frac{g^2}{2\pi^2}\Delta\ll D\ll \Delta$. Therefore, the logarithmic
term $\ln (D/\epsilon_0)$ in Eq.~(\ref{eq:ballistic_result}) is
approximately constant and of the order of $\ln(1/g)$. As the result,
for a grain in the ``ballistic'' regime, the contribution
(\ref{eq:ballistic_result}) yields a correction to the grain charge
\begin{equation}
\delta Q\sim e\frac{\ln(1/g)}{N_{\rm ch}} \frac{\delta{\cal N}}{{\cal
N}^*-{\cal N}} \label{eq:ball-corr}
\end{equation}
which is small outside the region ${\cal N}^*-{\cal N}\sim
\delta{\cal
  N}/N_\text{ch}$; this region is much smaller than $\delta{\cal N}$.

The ballistic estimate Eq.~(\ref{eq:ballistic_result}) holds if the
virtual excitation in the grain travels a distance shorter than the
grain size $L$ and the electron mean free path $\ell$; for
definiteness, we assume $L\lesssim\ell$. The length of the path of the
excitation depends on its typical energy given by $|E_0|$.  Indeed,
the velocity of excitation is $v\sim v_F\zeta/\Delta$, and the time of
travel is limited by $1/|E_0|$; here $v_F$ is the Fermi velocity. The
limitation on the excitation path length sets the condition for the
applicability of the ballistic approximation, $\Delta |E_0|\gtrsim
v_F^2/L^2$. The condition is violated sufficiently close to the charge
degeneracy point,
\begin{equation}
{\cal N}^*-{\cal N}\sim\frac{g v_F}{E_c L}. \label{eq:crossover}
\end{equation}
Here we assumed sufficiently large $N_{\rm ch}$, so that the interval
of ${\cal N}$ defined by Eq.~(\ref{eq:crossover}) is shorter than
$\delta{\cal N}$. We turn now to the estimate of the fourth-order
term (\ref{eq:fourth-order}) deep inside this interval.

If an excitation bounces off the grain walls many times, it is
reasonable to expect chaotization of its motion. This prompts one to
consider ensemble-averaged observables, rather than their specific
values for a given grain. To this end, we need to express $J$ of Eq.
(\ref{eq:fourth-order}) in terms of the correlation functions of the
electron states in the grain. We start with
representing~\cite{prada} the matrix elements $t_{kp}$ in terms of
the true electron eigenstates $\phi^*_k$ and $\chi_p$,
\begin{equation} \label{eq:tunnel-cont}
t_{kp}=\frac{\sqrt{\cal T}}{8 \pi^2 \nu} \int_S d^2\mathbf{x} \left.
\partial_z\phi^*_k(\mathbf{r})\partial_z\chi_p(\mathbf{r})\right|_{z=0}.
\end{equation}
Here, $\mathbf{x}$ is the longitudinal with respect to the barrier
component of coordinate $\mathbf{r}$. Inserting
Eq.~(\ref{eq:tunnel-cont}) into (\ref{eq:1.5}) and averaging
independently the states in the grain and in the lead, we can express
the conductance
\begin{eqnarray}
g =\frac{{\cal T}}{8 \pi^2\nu^2}\int_S d^2\mathbf{x}d^2\mathbf{x}'
\partial_z\partial_{z'}K_{\zeta\approx0}(\mathbf{r},\mathbf{r}')
\nonumber
\\
\times \left.\partial_z\partial_{z'}K_{\xi\approx
0}(\mathbf{r}',\mathbf{r})\right|_{z=z'=0},
\label{eq:conductance-GF}
\end{eqnarray}
in terms of the ensemble-averaged Green's function in the grain,
\begin{equation} \label{eq:spectral}
K_\zeta(\mathbf{r},\mathbf{r}')\equiv-\frac{1}{\pi}\mathrm{Im}
G^R_\zeta(\mathbf{r},\mathbf{r}')=\overline{\sum_k
\phi_k(\mathbf{r})\phi^*_k(\mathbf{r}')\delta (\zeta-\zeta_k)},
\end{equation}
and similarly in the lead $K_\xi\equiv(-1/\pi)\mathrm{Im}G^R_\xi$.
As the conductance is determined by tunneling events taking place on
the spatial range $1/k_F$ close to the junction, it is enough to
take the Green's functions for half-infinite spaces with the
appropriate boundary condition that they vanish at the interface.
Averaging each of them independently on the disorder in the metals,
we recover the result $g={\overline{\cal T}}N_\text{ch}$ of the
ballistic regime. At the same time, the correlation function
(\ref{eq:correlator}) is strongly affected by the
disorder.\cite{Hekking} Expressing Eq.~(\ref{eq:correlator}) in
terms of the Green's functions in the metals and ensemble-averaging
the product of two Green's functions, the authors of
Ref.~\onlinecite{Hekking} obtained
\begin{eqnarray}
K_\text{d}(\xi,\xi';\zeta,\zeta')=\frac{g^2}{128 \pi^5 \nu} \int
\frac{d^2\mathbf{x} d^2 \mathbf{x}'}{S^2} \left[
P_{\zeta'-\zeta}(\mathbf{r}',\mathbf{r})\right. \nonumber \\
\left. +P_{\zeta-\zeta'}(\mathbf{r},\mathbf{r}') \right]_{z=z'=0}.
\end{eqnarray}
Here $P_{\omega}(\mathbf{r},\mathbf{r}')$ describes the evolution of
the probability density to find an electron at a given point
$\mathbf{r}$. For diffusive motion (under the condition $\ell\ll
L$), the diffuson $P_{\omega}(\mathbf{r},\mathbf{r}')$ obeys the
diffusion equation,
\begin{equation} \label{eq:diff-eq}
\left( i \omega - D \mathbf{\nabla}_\mathbf{r}^2 \right)
P_{\omega}(\mathbf{r},\mathbf{r}')=\delta^3(\mathbf{r}-\mathbf{r}'),
\end{equation}
and $D=(v_F \ell )/3$ is the diffusion constant in the grain. At
frequencies $\omega$ smaller than the Thouless energy $D/L^2$, the
solution of Eq.~(\ref{eq:diff-eq}) reaches the universal
(zero-mode) limit where
\begin{equation} \label{eq:zero-mode}
P_{\omega}(\mathbf{r},\mathbf{r}')+P_{-\omega}(\mathbf{r}',\mathbf{r})\approx \frac{2
\pi}{V}\delta(\omega)
\end{equation}
is independent of the coordinates in the grain. In the case of a
smaller grain, $L\lesssim\ell$, Eq.~(\ref{eq:diff-eq}) does not
hold, but the universal limit for
$P_{\omega}(\mathbf{r},\mathbf{r}')$ is the same~\cite{ABG}, and is
reached at $\omega\lesssim v_F/L$. With the help of
Eq.~(\ref{eq:zero-mode}), we may now evaluate
Eq.~(\ref{eq:fourth-order-cont}) to find
\begin{equation} \label{eq:diffusive_result}
J_\text{d}\approx -\frac{g^2 \ln(1/g)}{64 \pi^3} \, \delta \,
\sqrt{\frac{2 \Delta}{-E_0}}.
\end{equation}
Here $\delta=1/\nu V$ is the electron level spacing in the grain.
Equation~(\ref{eq:diffusive_result}) yields (in the region ${\cal
  N}^*-{\cal N}\ll\delta{\cal N}$) a correction to the charge
of the order
\begin{equation}
\delta Q\sim e \frac{\ln(1/g)}{N_{\rm ch}^{\rm eff}}
               \left(\frac{\delta{\cal N}}
                          {{\cal N}^*-{\cal N}}\right)^{2},
\quad \frac{1}{N_{\rm ch}^{\rm eff}}=\frac{\delta}{g\Delta}
\label{eq:diff-corr}
\end{equation}
Comparing Eqs.~(\ref{eq:ball-corr}) and (\ref{eq:diff-corr}), we see
that they match each other at the gate voltage given by
Eq.~(\ref{eq:crossover}). The latter defines the crossover between
the ballistic and ``zero-mode'' limits, and corresponds to the path
length of the virtual excitations of the order of $L$. The
correction Eq.~(\ref{eq:diff-corr}) remains small everywhere except
a very narrow region around the charge degeneracy point, ${\cal
N}^*-{\cal
  N}\lesssim \delta {\cal N}/(N_{\rm ch}^{\rm eff})^{1/2}$.

On the odd side of the charge degeneracy point, at ${\cal N}>{\cal
  N}^*$, the unperturbed ground state is $P_o
\gamma^\dagger_{k\sigma}\left|e\right>$, where $k$ is the closest to
the Fermi level state. In the zeroth in $1/N_{\rm ch}^{\rm eff}$
limit, the ground state energy is zero. The correction induced in
  $E_0$ by the perturbation $V$ in the Hamiltonian $H_0$ is
found from equation
\begin{equation}
E_{0}=\sum_{p}' \frac{\left|\tilde{t}_{kp}\right|^2 \theta
(\xi_p)}{E_{0}-\epsilon_0-\xi_p}. \label{eq:excited-odd-0}
\end{equation}
As the tunneling matrix element involves a single state $k$, its
solution should exhibit strong mesoscopic fluctuations. The typical
value, however, is easy to find,
\begin{equation}
E_{0} \sim -\frac{1}{2\pi^2} g \delta \ln (1/g).
\label{eq:excited-odd}
\end{equation}
Therefore the coupling to the lead induces a small shift of ${\cal
  N}^*$ which scales proportionally to the one-level spacing in the
grain and is of the order of $g\delta/E_c$. Beyond this small shift,
the ground state energy at ${\cal N}>{\cal N}^*$ is not modified by
perturbation $V$.

Now we return briefly to the effect of the terms proportional to
$w_{kk^{\prime }}$ and $w_{pp^{\prime}}^i$ in $V$. Indeed, they do
also contribute to the Brillouin-Wigner expansion (\ref{eq:BW}) and
give correction to the ground state energy
(\ref{eq:ground-state-energy}). In particular, at ${\cal N}<{\cal
  N}^*$, the fourth-order-in-$H_t$ corrections to $E_0$ formed with
such terms are given by integrals similar to
Eq.~(\ref{eq:fourth-order-cont}). They only differ by the energy
ranges of integration which account for the virtual states outside
the bandwidth involved in the evaluation of $w_{kk^{\prime }}$ and
$w_{pp^{\prime}}^i$. The resulting correction is of the same order
as, and not more singular at ${\cal N}\to {\cal N}^*$ than the
contributions we have already evaluated, see
Eqs.~(\ref{eq:ballistic_result}) and (\ref{eq:diffusive_result}).

To summarize this Section, we have demonstrated that in the case of
a wide junction ($N_{\rm ch}\gg 1$) corrections to the ground state
energy Eq.~(\ref{eq:ground-state-energy}) are parametrically small.
The smearing of the non-analytical ${\cal N}$--dependence of the
grain charge, see Eq.~(\ref{eq:ground-state-charge}) vanishes in the
limit of zero level spacing $\delta$ in the grain. Identifying
$\left|0\right>$ and $\left|k\sigma\right>$ with
$P_e\left|\Phi\right>$ and
$P_o\gamma^\dagger_{k\sigma}\left|\Phi\right>$, respectively, where
$\left|\Phi\right>$ is the BCS ground state in the grain in the
grand-canonical ensemble, and discarding the terms proportional to
$w_{kk'}$ and $w_{pp'}^i$ in $V$, we can put Hamiltonian
(\ref{eq:effective-hamiltonian-2}) in the form
(\ref{eq:effective-hamiltonian}).

We note finally that in the opposite limit of a point contact, $N_{\rm
  ch}=1$, instead of the jump in $C_{\rm diff}$ we would find a smooth
crossover~\cite{Pesin} of width $\delta{\cal N}$ in the region of
even-odd transition.

\section{Charge steps at finite temperature} \label{sec:4}

In this Section, starting from the effective Hamiltonian
(\ref{eq:effective-hamiltonian}) for a wide, multichannel contact
between the superconducting grain and a normal lead, we derive
rigorously Eq.~(\ref{eq:Z}) which gives the partition function of
the system near the charge degeneracy point ${\cal N}^*$.

\subsection{Density of states}

To compute $Z$, we must determine the excited states.

Without coupling, the many-body ``even'' eigenstates of the system without
excitation in the grain are denoted $|0,f\rangle\equiv|0\rangle\otimes|f\rangle$.
We recall that $|0\rangle$ is the wavefunction for the grain in the even state,
while $|f\rangle$ is the wavefunction of the lead corresponding to a particular
set $f\equiv\{n_{p\sigma}\}$ of electron occupation numbers for each state
$(p\sigma)$. The energy of the state $|0,f\rangle$ is ${\cal E}_f+\epsilon_0$,
where
\begin{equation}
{\cal E}_f=\sum_{p\sigma}\xi_p\left(n_{p\sigma}-\theta(\xi_p)\right)
\end{equation}
is the excitation energy for the lead in state $|f\rangle$. The
many-body ``odd'' states with one excitation in state $(k\sigma)$ in
the grain are
$|k\sigma,f\rangle\equiv|k\sigma\rangle\otimes|f\rangle$, their
energy is ${\cal E}_f+\zeta^2_k/(2\Delta)$.

At finite coupling, even and odd state hybridize. In order to find
the eigenergies, we can still use Brillouin-Wigner perturbation
theory like we did for determining the ground state energy in
Section \ref{sec:3}. Starting from an unperturbed eigenstate
$|0,f\rangle$, we can write the Brillouin-Wigner equation for its
energy in the presence of the coupling by replacing
$\{|e\rangle,\epsilon_0,E_0 \}$ with $\{ |0,f\rangle,{\cal
E}_f+\epsilon_0,E \}$ in Eq.~(\ref{eq:BW}). The solutions of this
equation are the exact eigenenergies. For a wide junction (in the
limits $\delta \to 0$ and $N_\text{ch} \to \infty$), such equation
can still be truncated up to second order terms in the perturbation
$V$, like in Section~\ref{sec:3}. However the Brillouin-Wigner
equation generalizing Eq.~(\ref{eq:selfconst_energy_even})
\begin{equation}
E = \epsilon_0+{\cal E}_f +
    \sum_{kq\sigma}'
    \frac
        {\left|\tilde{t}_{kq}\right|^2 n_{q\sigma} }
        {E-{\cal E}_f+\xi_q-\zeta^2_k/(2\Delta)  }
, \label{eq:selfconst_excited}
\end{equation}
now defines a large number of excited states and is impractical to
solve for each of them. The prime in the sum means that only
unperturbed eigenstates in the bandwidth $D$ are included in the sum.
Here, let us note that Eq.~(\ref{eq:selfconst_energy_even}) is
obtained from (\ref{eq:selfconst_excited}) for the particular set
$f_0=\{n_{p\sigma}\}$ determined by the Fermi function at zero
temperature, $n_{p\sigma}=\theta(-\xi_p)$, and with the property
${\cal E}_{f_0}=0$ . At zero temperature, we were only looking for the
solution with lower energy, corresponding to the ground state.

The partition function (\ref{eq:Z-def}) which is the sum over the
full set of eigenstates $E_\lambda$ can be expressed in terms the
exact density of states, $\nu(E)$:
\begin{equation} \label{eq:Z-2}
Z=\int_{-\infty}^{\infty} d E \nu(E) e^{-\beta E},
\end{equation}
where $\nu(E)=\sum_\lambda\delta(E-E_\lambda)$. The density of state
is related to the exact Green's function $G_E=(E-H_0)^{-1}$ at
energy $E$:
\begin{equation}\label{eq:dos-1}
\nu(E)=-\frac{1}{\pi} \mathrm{Im}\, \mathrm{Tr}\, G_{E_+},\quad
E_+=E+i0^+.
\end{equation}
We evaluate $\mathrm{Tr}\,G$ in the basis of the unperturbed
eigenstates $|0,f\rangle$ and $|k\sigma,f\rangle$. Thus we can write
the density of states
\begin{equation}\label{eq:dos-2}
    \nu (E) =-\frac{1}{\pi} \mathrm{Im}\left[ \sum_f G_{E_+}(0,f) +\sum_{k\sigma,f}
    G_{E_+}(k\sigma,f)\right]
\end{equation}
in terms of the diagonal elements of $G_E$ in this basis:
\begin{subequations}
\begin{eqnarray}
G_E(0,f)&=&\langle 0,f|(E-H_0)^{-1}| 0,f \rangle, \\
G_E(k\sigma,f)&=&\langle k\sigma,f|(E-H_0)^{-1}| k\sigma,f \rangle.
\end{eqnarray}
\end{subequations}

In order to evaluate $G_E(0,f)$, we need to introduce the additional
matrix elements of $G_E$
\begin{subequations}
\begin{eqnarray}
G_E(0,f;0,f')&=&\langle 0,f|(E-H_0)^{-1}| 0,f' \rangle, \\
G_E(k\sigma,f;0,f')&=&\langle k\sigma,f|(E-H_0)^{-1}| 0,f' \rangle.
\end{eqnarray}
\end{subequations}
They solve the closed set of equations
\begin{subequations}
\begin{eqnarray}
\delta_{ff'}&=&(E-{\cal E}_f-\epsilon_0)G_E(0,f;0,f') \nonumber\\
&&-\sum_{kq\sigma}'\tilde{t}_{kq}n_{q\sigma}G_E(k\sigma,f_{q\sigma};0,f'),
\label{eq:evol-1}
\\
0&=&(E-{\cal E}_f-\zeta^2_k/(2\Delta))G_E(k\sigma,f;0,f')\nonumber\\
&&-\sum_{q}'\tilde{t}_{kq}^*(1-n_{q\sigma})G_E(0,f^{q\sigma};0,f'),
\label{eq:evol-2}
\end{eqnarray}
\end{subequations}
Here we defined $\delta_{ff'}=1$ if all the electron occupation
numbers in the sets $f$ and $f'$ coincides, otherwise
$\delta_{ff'}=0$. Moreover, the set $f_{q\sigma}$ (respectively
$f^{q\sigma}$) coincides with $f$, except for the occupation number
in state $(q\sigma)$ which is set to $n_{q\sigma}=0$ (respectively
$n_{q\sigma}=1$). Inserting (\ref{eq:evol-2}) into
(\ref{eq:evol-1}), and defining the self-energy
\begin{equation}  \label{eq:self-irreg}
\widetilde{\Sigma}_E(0,f)
=\sum_{kq\sigma}'\frac{|\tilde{t}_{kq}|^2n_{q\sigma}}{E-{\cal
E}_f+\xi_q-\zeta^2_k/(2\Delta)}
\end{equation}
we get the equation
\begin{eqnarray}
&&\delta_{ff'}=(E-{\cal E}_f-\epsilon_0-\widetilde{\Sigma}_E (0,f)) G_E(0,f;0,f')  \\
&&-\sum_{k,p\neq
q,\sigma}'\frac{\tilde{t}_{kq}\tilde{t}^*_{kp}n_{q\sigma}(1-n_{p\sigma})}
{E-{\cal
E}_f+\xi_q-\zeta^2_k/(2\Delta)}G_E(0,[f_{q\sigma}]^{p\sigma};0,f')
\nonumber
\end{eqnarray}
The second term in the r.h.s of this equation gives a negligible
contribution in the limit $N_\text{ch}\to\infty$. (This can be shown
along the same lines as in Section \ref{sec:3}.) As a result, we
find
\begin{equation} \label{eq:G0}
G_E(0,f)\equiv G_E(0,f;0,f)=[E-{\cal
E}_f-\epsilon_0-\widetilde{\Sigma}_E(0,f)]^{-1}.
\end{equation}
The last term in Eq.~(\ref{eq:ug}) defining $\epsilon_0$ depends on
bandwidth $D$. By subtracting this term from $\epsilon_0$ and adding
it to $\widetilde{\Sigma}_E(0,S)$, we write the Green's function
(\ref{eq:G0}) equivalently
\begin{equation}
  \label{eq:G_0_bis} G_E(0,f)= [E-{\cal E}_f-\omega_0-\Sigma_E(0,f)]^{-1},
\end{equation}
where $\omega_0=2E_c ({\cal N}-{\cal N}^*)$, and
\begin{equation} \label{eq:self-reg}
\Sigma_E(0,f)=\widetilde{\Sigma}_E(0,f)+\epsilon_0-\omega_0.
\end{equation}
Inserting Eqs.~(\ref{eq:ug}) and (\ref{eq:self-irreg}) into
(\ref{eq:self-reg}) we obtain
\begin{eqnarray}
    \Sigma_E(0,f) =   \sum_{kq\sigma}
    && \left\{ \frac
        {\left|\tilde{t}_{kq}\right|^2 n_{q\sigma} }
        {E-{\cal E}_f+\xi_q -\zeta^2_k/(2\Delta) }
\right. \nonumber \\
&& \left.
       -\frac
        {\left|\tilde{t}_{kq}\right|^2 \theta(-\xi_q) }
        { \xi_q -\zeta^2_k/(2\Delta)} \right\}. \label{eq:Sigma_0_bis}
\end{eqnarray}
At high energies $|\xi_p|$,  the electron occupation numbers
asymptotically behave as the zero temperature Fermi function.
Therefore, the sum in Eq.~(\ref{eq:Sigma_0_bis}) is convergent and
we don't need to specify anymore that only the states in the low
energy subspace are included in it.

We can determine $G_E(k\sigma,f)$ in a similar way. In the same
limit $N_\text{ch} \to \infty$, we obtain
\begin{eqnarray}
   G_E(k\sigma,f)&=&g_E(k\sigma,f)+g_E(k\sigma,f)^2  \nonumber\\
  && \times  \sum_p|\tilde{t}_{kp}|^2(1-n_{p\sigma})G_E(0;f^{p\sigma})
  \label{eq:Gk}.
\end{eqnarray}
The first term in the r.h.s. of Eq.(\ref{eq:Gk}) coincides with the
Green's function for the unperturbed state $|k\sigma,f\rangle $ with
one excitation in the grain, $g_E(k\sigma,f)=[E-{\cal
E}_f-\zeta^2_k/(2\Delta) ]^{-1}$. The second term comes from its
hybridization with the many-body states with no excitation in the
grain.

Inserting Eqs.~(\ref{eq:G_0_bis}) and (\ref{eq:Gk}) into
(\ref{eq:dos-2}), we express the density of states as the sum of
different contributions:
\begin{eqnarray}
\nu (E) &=&\sum_f\left[\sum_{k\sigma}\delta (E-{\cal
E}_f-\frac{\zeta^2_k}{2\Delta})+ \delta \nu_f (E)\right], \label{eq:dos-3}\\
\delta \nu_f (E) &=& -\frac{1}{\pi} \mathrm{Im}\left[ G_{E_+}(0,f)
\left( 1- \frac{d \Sigma_{E_+}(0,f)}{d E} \right)\right].
\label{eq:dos-4}
\end{eqnarray}
In Eq.~(\ref{eq:dos-3}), the sum of $\delta$-functions comes from
the first term in Eq.~(\ref{eq:Gk}) while the terms $\delta\nu_f(E)$
come from Eq.~(\ref{eq:G_0_bis}) and the second term in
Eq.~(\ref{eq:Gk}).

In the next Section, we use Eqs.~(\ref{eq:dos-3}) and
(\ref{eq:dos-4}) to evaluate the partition function.
We show that the contribution of $\delta \nu_f(E)$, Eq.~(\ref{eq:dos-4}), to
the partition function can be approximated by that of a single state with energy
$E =E_0+ {\cal E}_f$ and derive Eq.~(\ref{eq:Z}) for the partition function.
Finally, we show that Eqs.~(\ref{eq:Z}) and (\ref{eq:charge}) accurately account
for the thermodynamic quantities of the grain in the entire temperature range $T<
T^*$, and at any sign of ${\cal N}-{\cal N}^*$ as well.

\subsection{Partition function}

Inserting Eq.~(\ref{eq:dos-3}) into (\ref{eq:Z-2}), we obtain the
partition function
\begin{eqnarray}
Z  &=&   \sum_f
    e^{-\beta {\cal E}_f} \left(
        \sum_{k\sigma} e^{-(\beta\zeta^2_k/2\Delta)}
\right.
\nonumber
\\
    &&+ \left.    \int^{\infty}_{-\infty} d \omega e^{-\beta \omega}
    \delta\nu_f( \omega + {\cal E}_f )
    \right) .
\label{eq:Z-states}
\end{eqnarray}
Introducing the reduced partition function of the grain,
$\widetilde{Z}=Z/Z_\text{lead}$ and inserting
Eqs.~(\ref{eq:G_0_bis}) and (\ref{eq:self-reg}) into
(\ref{eq:dos-4}) we obtain
\begin{equation}\label{eq:partition_reduced}
    \widetilde{Z}  =  N_\text{eff}(T)
    -\int^{\infty}_{-\infty} d \omega \frac{e^{
    -\beta\omega}}{\pi} \mathrm{Im}
    \left\langle
    \frac{1 - d \Sigma_f(\omega_+)/d  \omega}{ \omega_+-\omega_0 -\Sigma_f(\omega_+)}
    \right\rangle_T.
\end{equation}
$Z_\text{lead}$ and $N_\text{eff}(T)$ were defined in
Eq.~(\ref{eq:Z}). In the second term of
Eq.~(\ref{eq:partition_reduced}), $\langle \ldots \rangle_T$ denotes
thermal averaging over with the Hamiltonian of the isolated lead and
we introduced the reduced self-energy
\begin{equation}
\Sigma_f(\omega)=\Sigma_{\omega+{\cal E}_f}(0,f).
\end{equation}

The integrand in the second term in Eq.~(\ref{eq:partition_reduced}) is
a regular function of the occupation numbers $n_{p\sigma}$. We can expand the
integrand in series in the set of $n_{p\sigma}$. Then we replace them by their
thermal average $f(\xi_p)=[1+\exp(\beta \xi_p)]^{-1}$ (Their fluctuations
$\langle [n_{p\sigma}-f(\xi_p)]^2\rangle_T$ scale as the inverse number of
electrons in the lead and can be safely ignored in the thermodynamic limit.)
Finally, we re-sum the series and obtain
\begin{equation}\label{eq:partition_average}
    \widetilde{Z}  =  N_\text{eff}(T)
    -\int^{\infty}_{-\infty} d \omega \frac{e^{
    -\beta\omega}}{\pi} \mathrm{Im}
    \frac{1 - d\Sigma(\omega_+)/d  \omega}{ \omega_+-\omega_0-\Sigma(\omega_+)},
\end{equation}
in terms of the thermally averaged self-energy
\begin{eqnarray}
    \Sigma(\omega)& \equiv & \langle \Sigma_f(\omega)\rangle_T
    \label{eq:Sigma_avergage} \\
    &=&
    \sum_{kq\sigma}
     \left\{ \frac
        {\left|\tilde{t}_{kq}\right|^2 f(\xi_q) }
        {\omega +\xi_q -\zeta^2_k/(2\Delta)}
       -\frac
        {\left|\tilde{t}_{kq}\right|^2 \theta(-\xi_q) }
        {\xi_q- \zeta^2_k/(2\Delta)} \right\}. \nonumber
\end{eqnarray}
In (\ref{eq:Sigma_avergage}) we can replace the summations over $k$
and $p$ by integrals, then integrate over $\zeta_k$, then take the
integrals over $\xi_p$ by parts. Finally, we get
\begin{equation}\label{eq:sigma_reduced}
    \Sigma(\omega_+) \equiv \frac{g}{16\pi T}\int^\infty_{-\infty} d \xi
     \frac{\sqrt{ 2\Delta  (-\omega_+ - \xi )}}{\cosh ^2 (\xi/2T) }.
\end{equation}

Equations (\ref{eq:partition_average}) and (\ref{eq:sigma_reduced})
express the thermodynamic quantities of the system in terms of two definite
integrals. They form the main results of this section. Below we show that the
grain charge may be approximated by Eq.~(\ref{eq:charge}).

If the Boltzmann weight $e^{-\beta \omega}$ is removed from the
integrand in the second term of Eq.~(\ref{eq:partition_average}) it
integrates to unity. Indeed, at $g=0$, this term corresponds to the
density of states of the single state of the grain without
quasiparticle on it. Therefore, by counting argument it must still
correspond to a single state at finite coupling. It is easy to see
that only the spectral weight of the Green's function residing at
frequencies $\omega \lesssim -\omega_i \equiv - T \ln
N_\text{eff}(T) $ results in non-negligible contribution to the
partition function in comparison with that of the first {\it large}
term. We can therefore restrict the integration over $\omega$ in the
second term of Eq.~(\ref{eq:partition_average}) to
$\omega<\omega_i$.

For $\omega < -\omega_i $ the integral in the last expression can be
readily evaluated and yields
\begin{equation}\label{eq:sigma_result}
    \Sigma(\omega_+) \approx \frac{g}{4\pi}\sqrt{2\Delta
    (-\omega)}\left[1+ {\cal O}\left(\frac{T}{|\omega|}\right)\right]
    - i \frac{g}{ 4  } \sqrt{\frac{ \Delta T}{
    2\pi  }}e^{\beta \omega}.
\end{equation}

It is clear from Eq.~(\ref{eq:sigma_result}) that the integral over
$\omega$ in Eq.~(\ref{eq:partition_average}) converges at the lower
limit. Its integrand has a pole precisely at $E_0<0$ when ${\cal
N}<{\cal N}^*$, see Eq.~(\ref{eq:ground-state-energy}). If $E_0 <
-\omega_i$ the pole $E_0$ lies within the region of integration over
$\omega$ and gives the dominant contribution to the integral which
equals to $\exp(-\beta E_0)$. At higher temperatures, the integral
in the second term in Eq.~(\ref{eq:partition_average}) may not be
evaluated by taking the residue at the pole. However at these
temperatures, the second term is negligible in comparison with the
first one. This is also true at ${\cal N}>{\cal N}^*$ when $E_0=0$,
and thus we obtain Eq.~(\ref{eq:Z}).

\section{Conclusion}\label{sec:5}

We studied charge quantization in a small superconducting grain with charging
energy $E_c>\Delta$ contacted by a normal-metal electrode in the limit
when the single particle mean level spacing in the grain, $\delta$, is small.
At zero conductance $g$ of the junction between the grain and the
electrode the steps in the charge {\it vs.} gate voltage dependence, $Q({\cal
N})$, are sharp and positioned in the dimensionless gate voltage ${\cal N}$ at
${\cal N}_0=2l \pm [1/2 +\Delta/(2E_C)]$.

At a finite conductance and zero temperature, the steps are shifted by
an amount $\propto g \Delta/E_C$, Eq.~(\ref{eq:shift}), making  the odd charge
plateaus even shorter. The charge steps become asymmetric and acquire a finite
width $W_0\propto g^2\Delta/E_C$.
We find the shape of the Coulomb blockade staircase, $Q({\cal N})$, for the
experimentally relevant case of a wide junction with a large number of tunneling
channels, $N_\text{ch}$. In the $N_\text{ch} \to\infty$ limit the ground state
energy of the system can be determined analytically and is given by
Eq.~(\ref{eq:ground-state-energy}). The resulting grain charge in the ground
state is given by Eq.~(\ref{eq:ground-state-charge}). Although the charge steps
are broadened at $g\neq 0$ and the dependence $Q({\cal N})$ becomes continuous,
the differential capacitance remains singular, displaying discontinuities at
certain values of the gate voltage.

At finite temperature $T<T^*= \Delta /\ln(\sqrt{8\pi}\Delta/\delta)$, we
obtain analytic expressions for the partition function in terms of two definite
integrals, Eqs.~(\ref{eq:partition_average}) and (\ref{eq:sigma_reduced}). The
resulting grain charge can be approximated by Eq.~(\ref{eq:charge}). The shape of
the charge step is plotted in Fig.~(\ref{fig:2}) for several characteristic
values of the system parameters. At temperatures exceeding the characteristic
``quantum'' temperature scale $T_q$, Eq.~(\ref{eq:T_q}), the steps acquire an
additional thermal shift described by Eq.~(\ref{eq:thermal-shift}). Finite
tunneling also leads to a non-monotonic temperature dependence of the steps width
with the minimal width achieved at $T\sim T_q$. The temperature dependent shift
of the step and its width are plotted in Fig.~\ref{fig:3}.

The existing experiment on such
systems\cite{Lafarge93b,Lafarge93a} perhaps is not sensitive
enough to see the quantum broadening evaluated in this paper.
Indeed one may think that the saturation of the ratio between even
and odd plateaus at low temperature observed in the experiment was
due to the quantum fluctuations rather than due to an impurity
creating a state within the gap, as suggested by the authors of
Ref.~\onlinecite{Lafarge93a}. This explanation however calls for
fairly large junction conductance of the order of $g=5$. This
value exceeds the experimental estimate of $25 \text{k}\Omega$ for
the resistance of the junction,\cite{LafargePhD} and also would
have lead to a significant slope of the charge plateaus, contrary
to the observations.\cite{Lafarge93a} Nevertheless, the evaluated
broadening for the N-I-S junction is easier to measure than the
charge step width in a normal-grain device (there the width is
exponentially small at small $g$). Recent experiment\cite{Lehnert}
mapped out a considerable portion of a charge step, broadened by
quantum fluctuations, in a normal device. We expect that similar
experiments with a hybrid device considered in this paper may
resolve the structure of entire step.

The authors are grateful to M. Pustilnik for valuable discussions.
The work at the University of Minnesota was supported by NSF grants
DMR-02-37296 and EIA-02-10736. The work of D.A.P. and A.V.A. was
supported by the NSF grant DMR-9984002, by the David and Lucille
Packard Foundation and by Lucent Technologies Bell Labs.

\end{document}